\documentclass[aps,prl,floatfix,twocolumn,superscriptaddress,showpacs,footinbib]{revtex4-1}
\usepackage{graphicx}% Include figure files
\usepackage{epstopdf}
\usepackage{bm}
\usepackage{color}

\graphicspath{{images/}}

\newcommand{\simlt}
      {\ifmmode       { \raisebox{-.8em}{$<$}\atop\sim}
         \else        {$\raisebox{-.8em}{$<$}\atop\sim$}
      \fi}

\begin{document}
\title{Atomic-scale coexistence of short-range magnetic order and superconductivity in Fe$_{1+y}$Se$_{0.1}$Te$_{0.9}$}

\author{Ramakrishna Aluru}
\affiliation{SUPA, School of Physics and Astronomy, University of St. Andrews, North Haugh, St. Andrews, Fife, KY16 9SS, United Kingdom}
\author{Haibiao Zhou}
\affiliation{SUPA, School of Physics and Astronomy, University of St. Andrews, North Haugh, St. Andrews, Fife, KY16 9SS, United Kingdom}
\author{Antoine Essig}
\affiliation{SUPA, School of Physics and Astronomy, University of St. Andrews, North Haugh, St. Andrews, Fife, KY16 9SS, United Kingdom}
\author{J.-Ph. Reid}
\affiliation{SUPA, School of Physics and Astronomy, University of St. Andrews, North Haugh, St. Andrews, Fife, KY16 9SS, United Kingdom}
\author{Vladimir Tsurkan}
\affiliation{Center for Electronic Correlations and Magnetism, Experimental Physics V, University of Augsburg, D-86159 Augsburg, Germany}
\affiliation{Institute of Applied Physics, Academy of Sciences of Moldova, MD 2028 Chisinau, Republic of Moldova}
\author{Alois Loidl}
\affiliation{Center for Electronic Correlations and Magnetism, Experimental Physics V, University of Augsburg, D-86159 Augsburg, Germany}
\author{Joachim Deisenhofer}
\affiliation{Center for Electronic Correlations and Magnetism, Experimental Physics V, University of Augsburg, D-86159 Augsburg, Germany}
\author{Peter Wahl}
\email{wahl@st-andrews.ac.uk}
\affiliation{SUPA, School of Physics and Astronomy, University of St. Andrews, North Haugh, St. Andrews, Fife, KY16 9SS, United Kingdom}

\date{\today}

\begin{abstract}
The ground state of the parent compounds of many high temperature superconductors is an antiferromagnetically (AFM) ordered phase, where superconductivity emerges when the AFM phase transition is suppressed by doping or application of pressure. This behaviour implies a close relation between the two orders. Understanding the interplay between them promises a better understanding of how the superconducting condensate forms from the AFM ordered background. Here we explore this relation in real space at the atomic scale using low temperature spin-polarized scanning tunneling microscopy (SP-STM) and spectroscopy. We investigate the transition from antiferromagnetically ordered $\mathrm{Fe}_{1+y}\mathrm{Te}$ via the spin glass phase in $\mathrm{Fe}_{1+y}\mathrm{Se}_{0.1}\mathrm{Te}_{0.9}$ to superconducting $\mathrm{Fe}_{1+y}\mathrm{Se}_{0.15}\mathrm{Te}_{0.85}$. In $\mathrm{Fe}_{1+y}\mathrm{Se}_{0.1}\mathrm{Te}_{0.9}$ we observe an atomic-scale coexistence of superconductivity and short-ranged bicollinear antiferromagnetic order.
\end{abstract}


\maketitle

% Motivation
The close proximity of magnetic order and superconductivity in the phase diagram of many unconventional superconductors such as cuprates, heavy fermion materials and iron-based superconductors\cite{dagotto_complexity_2005,uemura_superconductivity_2009, paglione_high-temperature_2010} suggests that understanding the relationship between these two phases might provide important clues towards an understanding of the physics of high temperature superconductivity. In the iron-based superconductors, both magnetic order and superconductivity (SC) originate from the iron $d$ bands. Quite a few of the iron-based superconductors exhibit regions in their phase diagrams, established from macroscopic characterization, where the two phases appear to coexist\cite{khasanov_coexistence_2009,goltz_microscopic_2014}, whereas in others there is clear phase separation\cite{luetkens_electronic_2009}.
Real-space imaging by spin-polarized STM can reveal whether superconductivity and magnetism really coexist at the atomic scale, or reside in spatially separated regions which are either antiferromagnetically ordered or superconducting.
From theory, it has been shown that in the iron-pnictides,  antiferromagnetic (AFM) order and superconductivity (SC) can coexist if the superconducting order parameter is of the sign-changing $s^\pm$-type , whereas for non-sign changing $s$-wave superconductivity, the two phases compete and superconductivity is suppressed once magnetic order sets in\cite{PhysRevB.81.174538,Fernandes2010,drew2009coexistence}.

Here, we study the relation between antiferromagnetic order and superconductivity in the iron chalcogenides, Fe$_{1+y}$Se$_{x}$Te$_{1-x}$, using low temperature spin-polarized scanning tunneling microscopy and spectroscopy (SP-STM/STS). The iron chalcogenides have the simplest crystal structure among the different classes of iron based superconductors. The crystal structure consists of layers of iron atoms forming a square lattice with selenium and tellurium atoms centered above or below these squares. The iron chalcogenides exhibit a natural cleavage plane which exposes a non-polar surface. The non-superconducting parent compound Fe$_{1+y}$Te with y$\leq0.1$ exhibits long-range commensurate magnetic order \cite{Li2009,Enayat2014} with a wave vector $\mathbf{q}_\mathrm{AFM}=(\pi,0)$.

\begin{figure}[!tbp]
\includegraphics[width=\columnwidth]{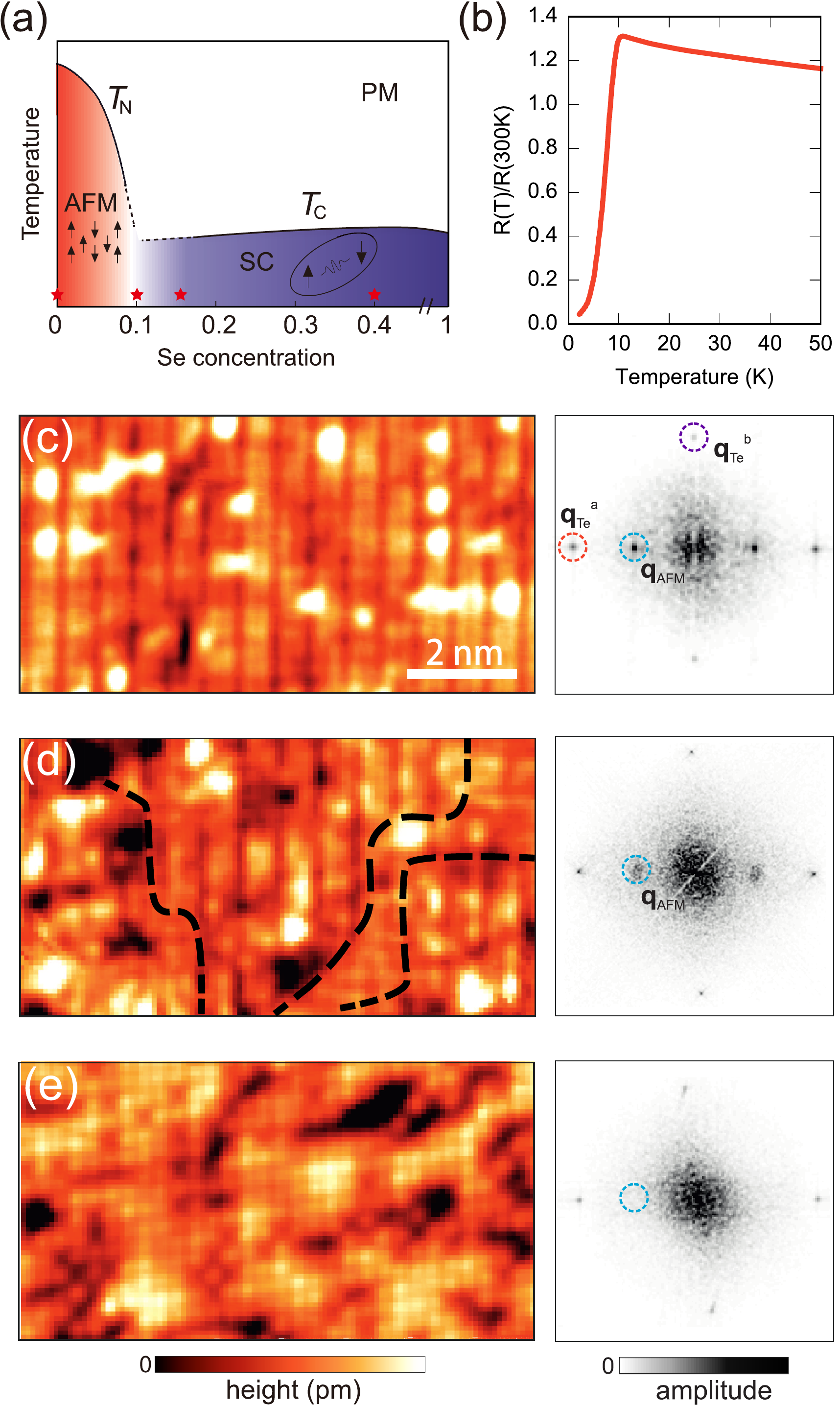}
\caption{\textbf{Magnetic order in Fe$_{1+y}$Se$_{x}$Te$_{1-x}$.} (a) Phase diagram of Fe$_{1+y}$Se$_{x}$Te$_{1-x}$ as a function of selenium (Se) concentration $x$ and temperature. The region between the antiferromagnetic (AFM) order and superconductivity, where the two phases appear to coexist, has also been described as a spin-glass phase\cite{Katayama2010}. PM: paramagnetic metal. (b) Temperature dependence of the resistivity of the sample with $x = 0.1$ used for STM measurements, showing an abrupt drop at $T_{\mathrm C} \sim$ 10 K due to the onset of superconductivity. (c-e) Topographic images obtained with magnetic tips on the samples with $x=0$ (c), $0.1$ (d) and $0.15$ (e). The topographies show the evolution from long-range magnetic order in (c) to short-range patches in (d) and suppression of the magnetic order in (e). The dashed lines in (d) indicate the boundaries between the regions with and without magnetic order. Panels to the right of (c-e) show the intensity of the Fourier tranform with the peaks associated with the Te/Se lattice and magnetic order highlighted (right panels).}
\label{Figure_1}
\end{figure}

With Se doping the magnetic order becomes short-ranged and incommensurate \cite{bao2009} and superconductivity sets in \cite{Mizuguchi2010,Katayama2010}. At higher Se concentration, x $\geq$ 0.4, bulk superconductivity is achieved \cite{Katayama2010,udai2013} (compare fig.~\ref{Figure_1}a). Despite a different magnetic ordering wave vector between the non-superconducting parent compounds of the iron chalcogenides and pnictides both show enhanced spin fluctuations at $\mathbf{q}=(\pi,\pi)$ in the superconducting state\cite{Liu2010,dai_antiferromagnetic_2015}.

In this work, we study the transition from antiferromagnetic order at $x=0$ to superconductivity at $x=0.15$ in Fe$_{1+y}$Te$_{1-x}$Se$_x$ by real space imaging of both the magnetic order and the superconducting pairing. We especially focus on a region of the phase diagram (Fig.~\ref{Figure_1}a) near $x=0.1$ where signatures of both, magnetic order and superconductivity (Fig.~\ref{Figure_1}b), are observed\cite{khasanov_coexistence_2009}.
% \cite{Khasanov2009}
SP-STM/STS measurements were carried out using a home-built low-temperature STM operating at 1.7 K which allows for \emph{in situ} sample transfer and cleavage \cite{white2011}. STM tips have been cut from a platinum-iridium wire and prepared by field emission on a gold single crystal. Magnetic tips have been prepared either by picking up interstitial excess iron atoms from the surface of the material or by gentle indentation of the tip into the sample surface \cite{Singh2015,Enayat2014}. In this work, we have prepared magnetic tips also by applying voltage pulses with amplitude of about 2 V which render the tips magnetic and remove excess iron atoms underneath the tip from the sample surface. STM images shown for the sample with $x=0.1$ have been obtained in areas where the excess iron atoms had been removed, as otherwise the magnetic contrast was obscured by the high density of excess iron atoms.

Topographic images were recorded in constant current mode. The differential tunneling conductance $\frac{dI}{dV}(V)$ was measured through a lock-in amplifier with $f=413~\mathrm{Hz}$ and modulation amplitude $V_\mathrm{mod}=800~\mathrm{\mu V}$. The bias voltage is applied to the sample.

The single crystals of Fe$_{1+y}$Se$_{x}$Te$_{1-x}$ ($x$ = 0, 0.1 and 0.15) were grown by the self-flux method \cite{gnezdilov_anomalous_2011,tsurkan_physical_2011}. The compositions of the samples as determined from energy-dispersive x-ray (EDX) analysis are Fe$_{1.07}$Te, Fe$_{1.05}$Te$_{0.89}$Se$_{0.11}$, Fe$_{1.08}$Te$_{0.85}$Se$_{0.15}$, respectively.

%Fermi surface schematic showing a hole like band centered at \textbf{k} = (0,0) and an electron like band centered at \textbf{k} = ($\pi,\pi$) in Fe$_{1+y}$Se$_{0.1}$Te$_{0.9}$. There exists no Fermi surface pocket which can be nested by $\mathbf Q_\mathrm{AFM}$ = ($\pi/2,\pi/2$). The superconducting spin resonance occurring around ($\pi,\pi$) is denoted by $\mathbf Q_\mathrm{SC}$.

The evolution of the magnetic order as function of Se doping for the samples with $x = 0$, $0.1$ and $0.15$ is shown in topographic images in Fig.~\ref{Figure_1}(c-e), together with the intensity of the Fourier transformation. The Fourier transformations show peaks associated with the surface atomic lattice of tellurium/selenium at $\mathrm{q}_\mathrm{Te}^a$, $\mathrm{q}_\mathrm{Te}^b$ and of the magnetic order at $\mathrm{q}_\mathrm{AFM}^a$. The long range unidirectional AFM order can be clearly seen by sharp peaks at $\mathrm{q}_\mathrm{AFM}^a$ in Fe$_{1+y}$Te (Fig.~\ref{Figure_1}c), while the magnetic order becomes short-range in the $x = 0.1$ sample (Fig.~\ref{Figure_1}d), evidenced by patches of magnetic order in the real space image (see black lines demarking magnetically ordered areas in Fig.~\ref{Figure_1}d) and significant broadening of the peak associated with the magnetic order in Fourier space. A resistivity measurement performed on the same sample used for STM shows that the sample exhibits a superconducting transition occurs at $T = 10 \mathrm{K}$, evidenced by the sharp drop in resistivity (Fig.~\ref{Figure_1}b). However, the transition is broader than in samples with larger $x$ and the resistivity does not go to zero, implying that the superconductivity is filamentary. At $x=0.15$, no magnetic contrast is detected (Fig.~\ref{Figure_1}e). From this series of topographic images, we can clearly see that the $x = 0.1$ sample lies in a phase with short-range magnetic order, while in the sample with $x = 0.15$, the magnetic order is already completely suppressed.

\begin{figure}[htbp]
\includegraphics[width=0.95\columnwidth]{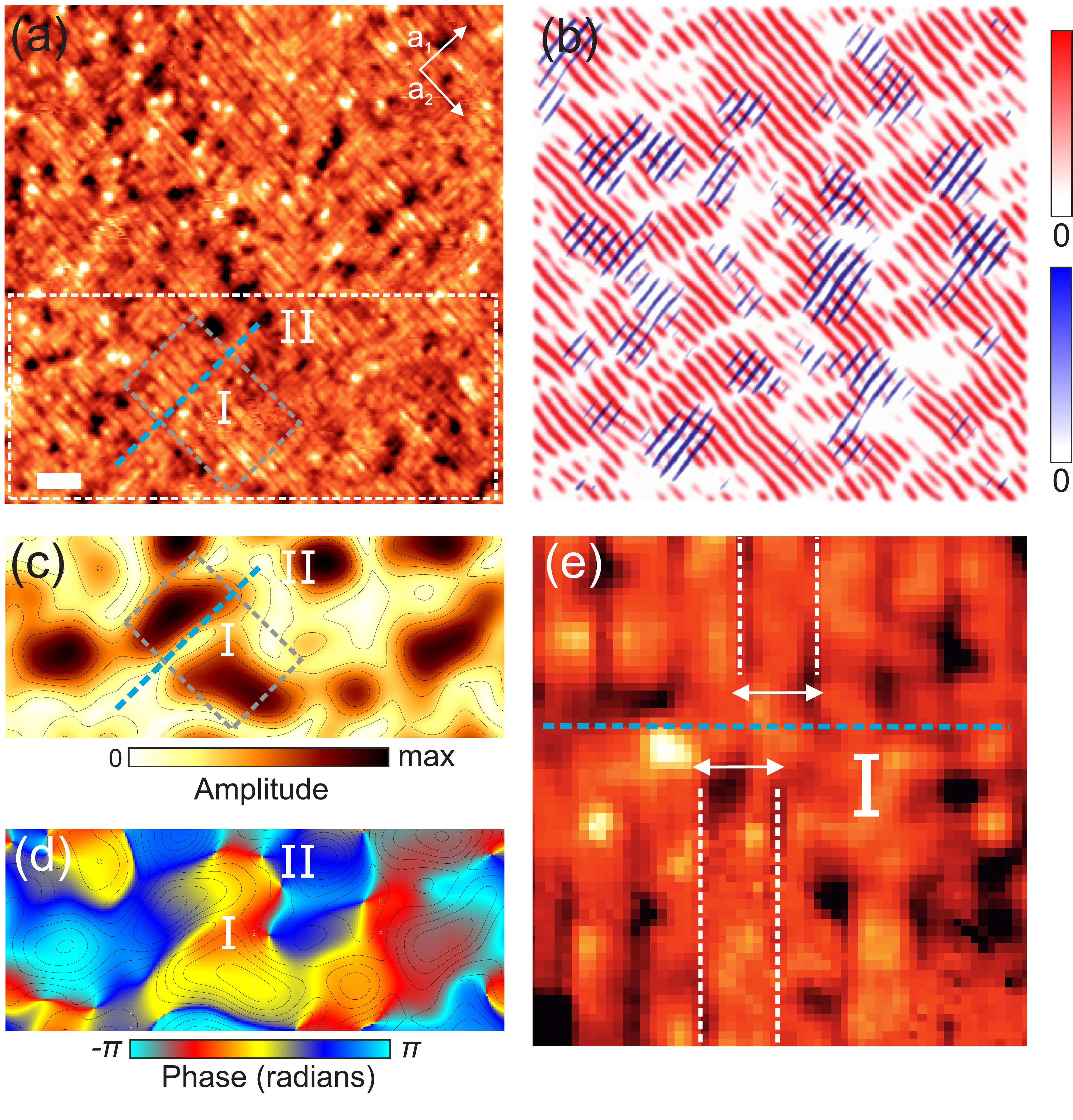}
\caption{\textbf{Short-range AFM order at $x=0.1$.} (a) Topography acquired with a magnetic tip which shows stripes in both crystallographic directions, $a_1$ and $a_2$. ($V_s=-50$ mV, $I_s=300$ pA, $30\times30$ nm$^{2}$). (b) Map obtained by overlaying the Fourier filtered images associated with magnetic order along $a_1$ (red) and $a_2$ (blue). The two Fourier-filtered images are encoded in different colors.  (c) Map showing the intensity of the magnetic order along the $a_1$ direction extracted from (a) using a lock-in analysis, the area shown corresponds to the white framed area in (a). (d) Phase map extracted from the lock-in analysis. (e) Magnetic order in the region indicated by a dashed rectangle in (a) and (c) to highlight the phase shift between the magnetic order in different regions of the sample.}
\label{Figure_2}
\end{figure}

The magnetic order is spatially inhomogeneous, and we can identify regions showing patches of magnetic order in both crystallographic directions, as seen in the topography in Fig.~\ref{Figure_2}a. This is clearly in contrast with the non-superconducting parent compound, where at low excess iron concentration $y$ the magnetic order is unidirectional, stablized by the anisotropy of the crystal structure.
To highlight the magnetically ordered patches, we show in Fig.~\ref{Figure_2}b a Fourier filtered image, obtained by overlaying the Fourier filtered images associated with the magnetic peaks at $\mathrm{q}_\mathrm{AFM}^{a_1}$ and $\mathrm{q}_\mathrm{AFM}^{a_2}$.
%(see details in Supplementary Fig).
Unlike Fe$_{1+y}$Te at high excess iron concentrations $\mathrm{y} \geq 0.12$, where by spin-polarized STM a bidirectional magnetic order is observed over extended surface regions \cite{Enayat2014}, here the magnetic order stays largely unidirectional within localized patches (Fig.~\ref{Figure_2}b). These observations raise two questions: (1) is it only the spin-polarization at the Fermi level which is spatially modulated, leading to a patchy appearance in STM images while the magnetic order in reality has long-range coherence? (2) If it is true short range order, what is the characteristic length scale ?
(1) can be tested by analyzing the phase of the magnetic order: if it is only the intensity which is modulated but the order long-ranged, spatially separated patches of magnetic order should exhibit the same phase - whereas if the magnetic order is localized in separate regions of the sample, the phase will be random between spatially separated patches. To distinguish these two scenarios, we have analyzed the local phase of the magnetic order from a lock-in analysis\cite{slezak_imaging_2008,lawler_intra-unit-cell_2010}.

\begin{figure}[htbp]
\includegraphics[width=\columnwidth]{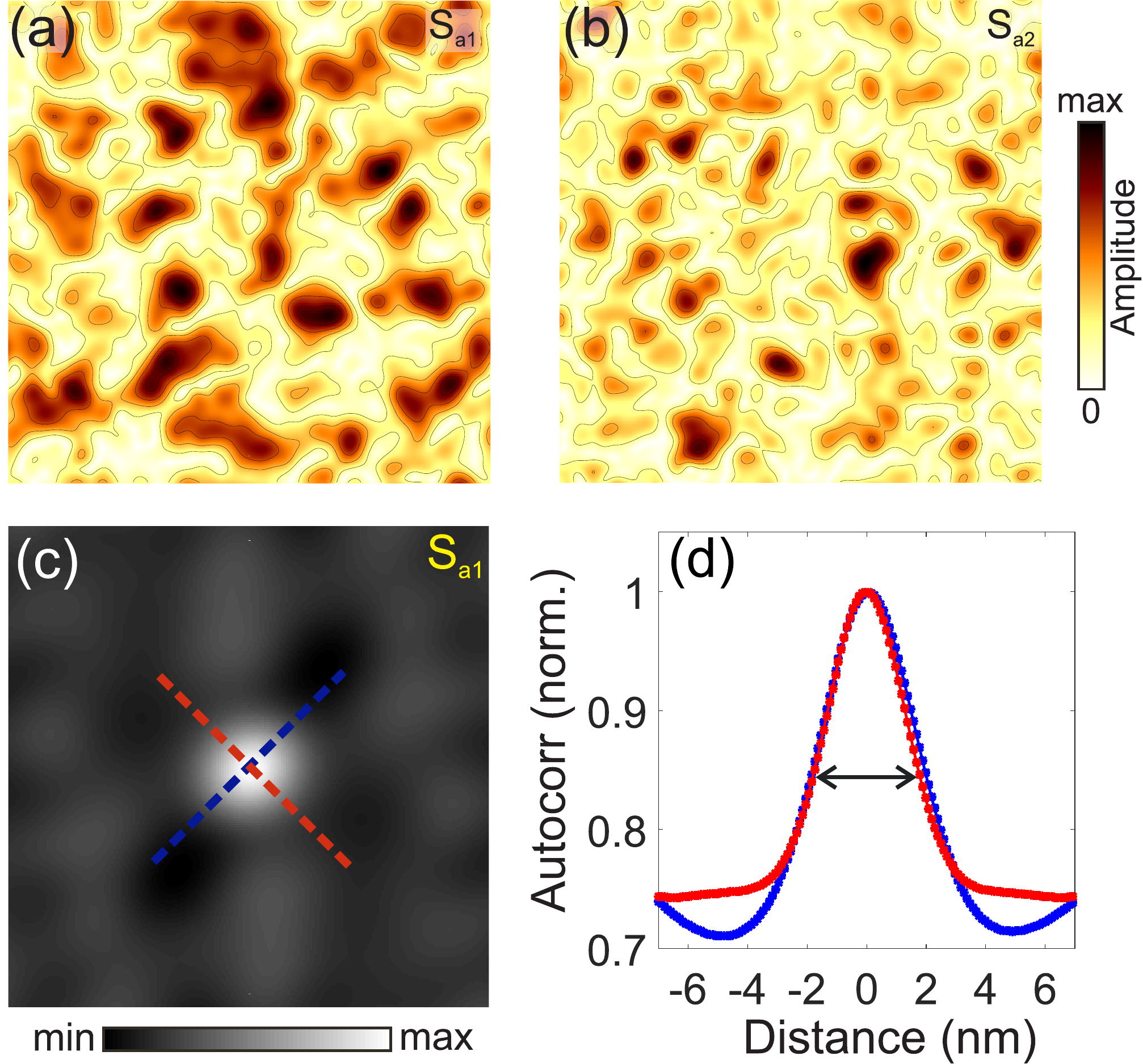}
\caption{\textbf{Correlation length of magnetic order in Fe$_{1+y}$Se$_{0.1}$Te$_{0.9}$.} (a-b) Maps of the intensity of stripes along $a_1$ and $a_2$, respectively, extracted from Fig.~\ref{Figure_2}a using a lock-in analysis. (c) Autocorrelation of the stripe intensity map along $a_1$ as shown in (a). (d) Line-cuts across the center of the autocorrelation images along both axes to extract the correlation length of the magnetic order (colours refer to cuts in (c)). The full width of the autocorrelation yields a characteristic length scale of $4.0\mathrm{nm}$ and $3.4\mathrm{nm}$ for the maps in (a) and (b), respectively.}
\label{Figure_3}
\end{figure}

In Fig.~\ref{Figure_2}(c, d), we show the results of a lock-in analysis of the topography in Fig.~\ref{Figure_2}(a), with the amplitude shown in (c) and the corresponding phase map in (d). As can be seen by comparison of Fig.~\ref{Figure_2}c and d, different regions with large intensity of the magnetic order exhibit different phase (compare, e.g., regions marked I and II). The analysis thus confirms that the sample really exhibits patches with magnetic order. Fig.~\ref{Figure_2}e shows an area within the dotted rectangle in Fig.~\ref{Figure_2}a showing the
change of the phase of the magnetic order between two different neighbouring regions, confirming the result from the lock-in analysis.
The same lock-in analysis can be used to address point (2), the characteristic length scale of the magnetically ordered patches. Fig.~\ref{Figure_3}a and Fig.~\ref{Figure_3}b show the extracted intensity maps S$_{a_1}$ and S$_{a_2}$. The characteristic size of the patches estimated from the autocorrelation, as shown in Fig.~\ref{Figure_3}c and d for S$_{a_1}$ is about 4 nm, similar to what an analysis of the width of the Fourier peak (compare fig.~\ref{Figure_1}d) yields. Analysis of the magnetic modulation for $a_2$ yields a similar though slightly smaller characteristic length scale. The extracted length scale from our SP-STM measurements is slightly larger than the value of $2.4\mathrm{nm}$ reported for selenium concentration $x=0.1$ from neutron scattering\cite{Katayama2010}.

\begin{figure}[htbp]
\includegraphics[width=0.95\columnwidth]{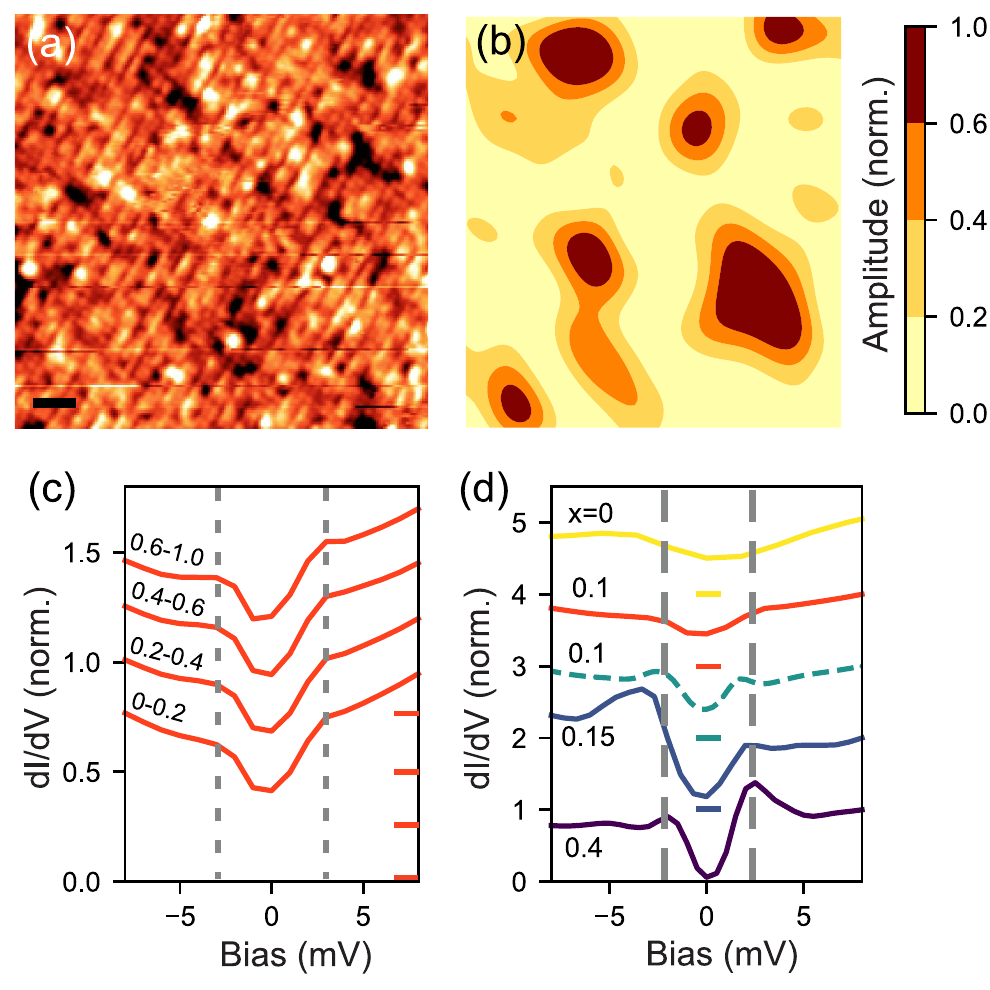}
\caption{\textbf{Relation between magnetic order and superconductivity in Fe$_{1+y}$Se$_{0.1}$Te$_{0.9}$.} (a) SP-STM topography showing magnetic and non-magnetic regions on the surface. The size is 19$\times$19 nm$^2$ and the scale bar is 2 nm. (b) Map showing the intensity of the magnetic order. (c) Spatially averaged spectra obtained from regions in (b) with similar strengths of magnetic order. All spectra are normalized at $V=8\mathrm{mV}$. (d) A comparison of the spectra in samples with $x$ = 0, 0.1, 0.15 and 0.4. Two spectra are shown for the sample with $x=0.1$. The red one is for the area shown in (a) and the dashed one is for the area shown in Fig.S2. Spectra are offset vertically for clarity in (c) and (d), a horizontal bar indicates the offset for each spectrum.}
\label{Figure_4}
\end{figure}

In order to address the relation between magnetic order and superconductivity, we have acquired spectroscopic maps consisting of tunneling spectra taken in the energy range of the superconducting gap and covering the same area in which we have characterized the magnetic order using spin-polarized STM. Fig.~\ref{Figure_4}a shows the SP-STM image of the region where the map has been taken. The map of strength of magnetic order extracted from the same image is shown in Fig.~\ref{Figure_4}b.
To analyze the relation between superconductivity and magnetic order, we have averaged the tunneling spectra in regions with similar strength of the magnetic order, shown in Fig.~\ref{Figure_4}c. A clear gap can be seen at the Fermi energy, which shows only negligible variation between regions with different strengths of the magnetic order. While the absence of a clear correlation is true for different regions of the sample, the overall shape of the spectra shows significant variation (compare Fig.~\ref{Figure_4} and Fig.S2), which we attribute to variations in the concentration of excess iron atoms. Comparison with the gap size seen at other Se concentrations, in particular the one in bulk superconducting $\mathrm{FeSe}_{0.4}\mathrm{Te}_{0.6}$\cite{udai2013} (see Fig.~\ref{Figure_4}(d)), shows that the gap size is quite similar, while the coherence peaks are less pronounced in the sample with $x=0.1$.

The absence of correlation between the antiferromagnetic order and superconductivity in our experiment puts a stringent constraint on the theories aimed at explaining superconductivity in iron-based materials.
Our results are consistent with the expectation for $s_\pm$ superconductivity, where the magnetic order only has a rather minor impact on the strength of the superconducting pairing, even though it would have an impact on the spin fluctuations. The latter will however remain rather localized in $q$-space to the wave vectors of the magnetic order. At the same time, it would be difficult to reconcile our experiments with conventional isotropic $s$-wave superconductivity\cite{vavilov_coexistence_2010, fernandes_competing_2010}, as would be promoted by orbital fluctuations \cite{kontani_orbital-fluctuation-mediated_2010}.
That the gap size which we observe remains essential constant for different selenium concentrations $x$ agrees with the fact that in FeSe$_{x}$Te$_{1-x}$, superconductivity does not disappear with decreasing $x$ due to a suppression of the transition temperature but rather due to the superconductivity becoming filamentary. This, together with observation of coexistence between magnetism and superconductivity in thin films of FeTe\cite{manna_interfacial_2017}, suggests that it is the excess iron which kills superconductivity for $x\rightarrow 0$ in FeSe$_{x}$Te$_{1-x}$, rather than the magnetic order. The lower excess iron concentration seen at the surface compared to the bulk may mean that the superconductivity in the bulk is not as prevalent as our measurements suggest, yet this does not impact on the principal finding of superconductivity coexisting at the atomic scale with antiferromagnetic order. The characteristic length of the magnetic order which we find is surprisingly consistent with neutron scattering\cite{Katayama2010}, whereas the ordering wave vector is not: from neutron scattering, the magnetic order becomes incommensurate with increasing Selenium concentration, which we do not observe. This may again be related to the different excess iron concentration at the surface.

Our results show atomic scale images of the coexistence of antiferromagnetic order and superconductivity for an iron chalcogenide material. They reveal a local coexistence of the two phases, with little impact of the  antiferromagnetic order on the superconductivity. Given the different magnetic ordering wave vectors in the iron chalcogenides compared to the iron pnictides, it remains to be seen whether the same holds true for the iron pnictide superconductors.

\begin{acknowledgments}
HZ, JPR and PW acknowledge funding from EPSRC through EP/I031014/1. VT, AL and JD acknowledge funding from the Deutsche Forschungsgemeinschaft (DFG) via the Transregional Collaborative Research Center TRR 80 (Augsburg, Munich, Stuttgart). Underpinning data will be made available at DOI:10.17630/129937d5-f15c-4ccd-ab0e-e407740ef4a2.
\end{acknowledgments}

%\bibliographystyle{apsrev4-1}
%\label{Bibliography}
%\bibliography{fesete_bib}

\section*{Supplementary Information}

\begin{figure}[htbp]
\includegraphics[width=0.65\textwidth]
{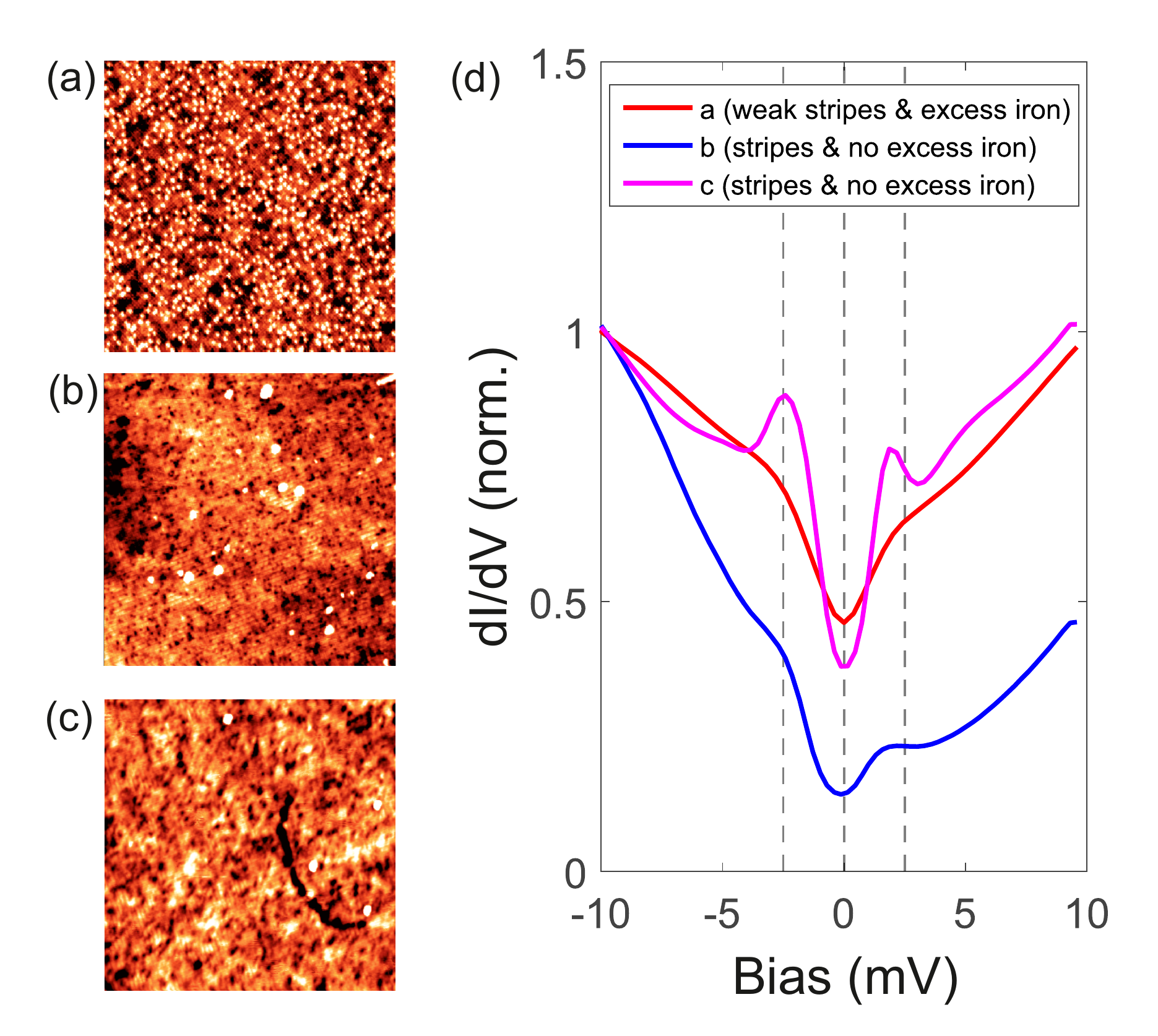}
\caption{\textbf{Variability of tunneling spectra in $\mathrm{Fe}_{1.05}\mathrm{Se}_{0.1}\mathrm{Te}_{0.9}$.} (a) STM topography showing weak magnetic order and excess iron atoms. (b-c) STM topography showing regions with magnetic order and no excess iron atoms at the surface. (d) Average spectra obtained from the regions shown in (a-c). ($V_s=50$ mV, $I_s=400$ pA, $V_\mathrm{mod}=800$ $\mathrm{\mu V}$). All the spectra are normalized at $-10\mathrm{mV}$. While the spectra vary substantially in shape, the gap size is consistent between all of them.}
\label{STS_exFe}
\end{figure}

%\subsection{Another data set}

\begin{figure}[htbp]
\includegraphics[width=0.65\columnwidth]
{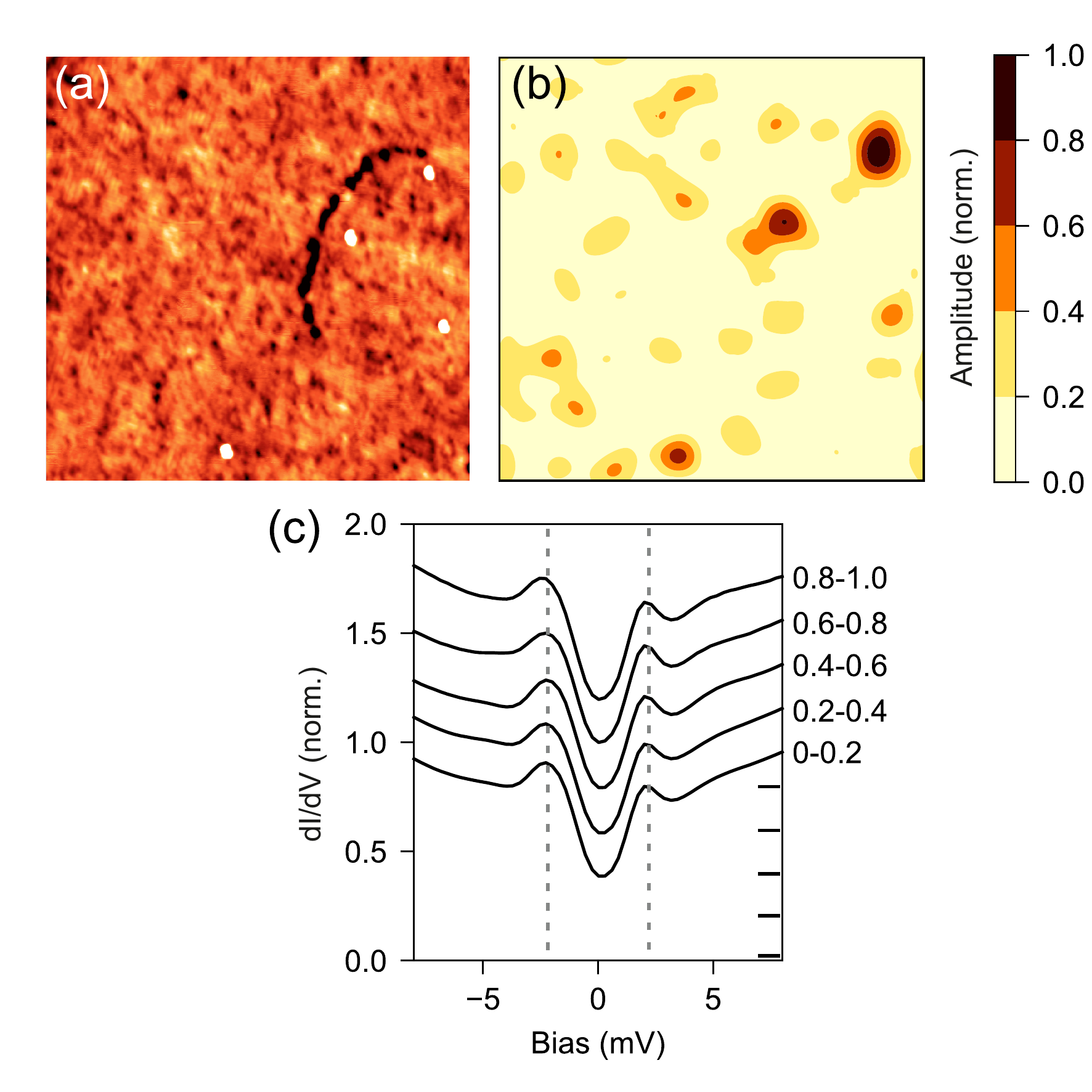}
\caption{\textbf{Coexistence of AFM and SC in another area.} (a) Topographic image of a different area of the  $\mathrm{Fe}_{1.05}\mathrm{Se}_{0.1}\mathrm{Te}_{0.9}$ sample than the one shown in Fig. 4, showing short-range magnetic order. ($V_s=80$ mV, $I_s=100$ pA, $50\times50$ nm$^{2}$). (b) Map of the amplitude of the magnetic order obtained from (a). (c) Tunneling spectra ($V_s=50$ mV, $I_s=400$ pA, $V_\mathrm{mod}=800$ $\mathrm{\mu V}$) extracted from a spectroscopic map taken in the same region as (a) and averaged in regions with the same intensity of magnetic order. The results are fully consistent with the ones shown in Fig. 4 in the main text.}
\label{STS_region2}
\end{figure}

\end{document}